\documentclass[usenatbib,usegraphicx]{mn2e}

\begin{document}

\title{Heated Disc Stars in the Stellar Halo}

\author[C. W. Purcell et al.]
{Chris W. Purcell,$^1$\thanks{E-mail: cpurcell@uci.edu}
James S. Bullock,$^1$
and Stelios Kazantzidis$^2$ \\
$^1$Center for Cosmology, Department of Physics and Astronomy, The University of California, Irvine, CA 92697 USA \\
$^2$Center for Cosmology and Astro-Particle Physics; and Department of Physics; and Department of Astronomy, \\
The Ohio State University, Columbus, OH 43210 USA
}
\date{Accepted 2010 January 23.  Received 2010 January 6; in original form 2009 November 17}

\maketitle

\begin{abstract}  

Minor accretion events with mass ratio $M_{\rm sat}$:$M_{\rm host} \simeq$~1:10 are common in the context of 
$\Lambda$CDM cosmology.  We use high-resolution simulations of Galaxy-analogue systems to show that these 
mergers can dynamically eject disc stars into a diffuse light component that resembles a stellar halo both spatially 
and kinematically.  For a variety of initial orbital configurations, we find that $\sim 3-5 \times 10^8 \, M_\odot$ of 
primary stellar disc material is ejected to a distance larger than 5 kpc above the galactic plane.  
This ejected contribution is similar to the mass contributed by the tidal disruption of the satellite galaxy itself, 
though it is less extended.  If we restrict our analysis to the approximate solar neighborhood in the disc plane, 
we find that $\sim 1\%$ of the initial disc stars in that region would be classified kinematically as halo stars.  Our 
results suggest that the inner parts of galactic stellar halos contain ancient disc stars and that these stars may have 
been liberated in the very same events that delivered material to the outer stellar halo.

\end{abstract}

\begin{keywords}
Cosmology: theory --- galaxies: formation --- galaxies: evolution
\end{keywords}

\section{Introduction} 

The canonical paradigm describing the emergence of outer stellar halos around galaxies involves the accretion and tidal disruption of 
satellite systems \citep{Searle_Zinn78,Johnston_96,Cote_etal00,Bullock_01,Bullock_Johnston05,Abadi_etal06,Delucia_08,Johnston_08,Cooper_09}. 
Empirical evidence suggests that these predicted mergers have indeed populated the outskirts of the Milky Way and M31 halos, 
with diffuse, stream-like signatures apparent at low surface brightness 
\citep[recently,][and references therein]{McConnachie_09,Gilbert_09a,Gilbert_09b,Starkenburg_09,Bell_08}.

Perhaps unsurprisingly, significant diffuse stellar halo components also appear to be pervasive in the population of nearby disc 
galaxies outside the Local Group \citep{Sackett_etal94,Morrison_etal97,Lequeux_etal98,Abe_etal99,Zibetti_Ferguson04}.  
The GHOSTS survey described by \citet{deJong_etal07a} has established the presence of relatively luminous and extended
halos surrounding the massive disc galaxies NGC253, NGC891, and M94; the $V$-band surface brightness $\mu_V$ of 
this material is typically $\sim 28-29$~mag arcsec$^{-2}$ at large heights ($\sim 20-30$~kpc) above the disc plane.  This 
quantity of diffuse light cannot easily be reconciled with a model where most of the inner, richer stellar halo is built by the destruction 
of dwarf satellite galaxies accreted at early times, since these old and faint subhalos can add {\em at most} only a few 
percent to the total luminosity of the host galaxy and produce metal-poor stellar halos \citep{Purcell_etal07,Purcell_etal08}.  
This accreted light is typically deposited over such a large 
volume that the resulting ancient halo's surface brightness is far below that observed in the less-extended spheroids enveloping 
many massive disc galaxies.  Faint stellar halo material around M33 \citep[][see also \citealt{Chapman_etal06,Kalirai_etal06,Helmi08}]{McConnachie_etal06} 
and diffuse extended light around dwarf galaxies like the LMC \citep{Minniti_etal03} are even more difficult to explain in the 
accretion scenario \citep{Purcell_etal07}.  These general arguments, along with direct evidence that the inner halo of M31 
\citep{Guha_etal05} and the Milky Way \citep{Carollo_etal07} are populated by at least two distinct components, make it clear 
that a multivalent picture of inner stellar halo formation must emerge.


\begin{figure*}
\includegraphics[width=2.1in]{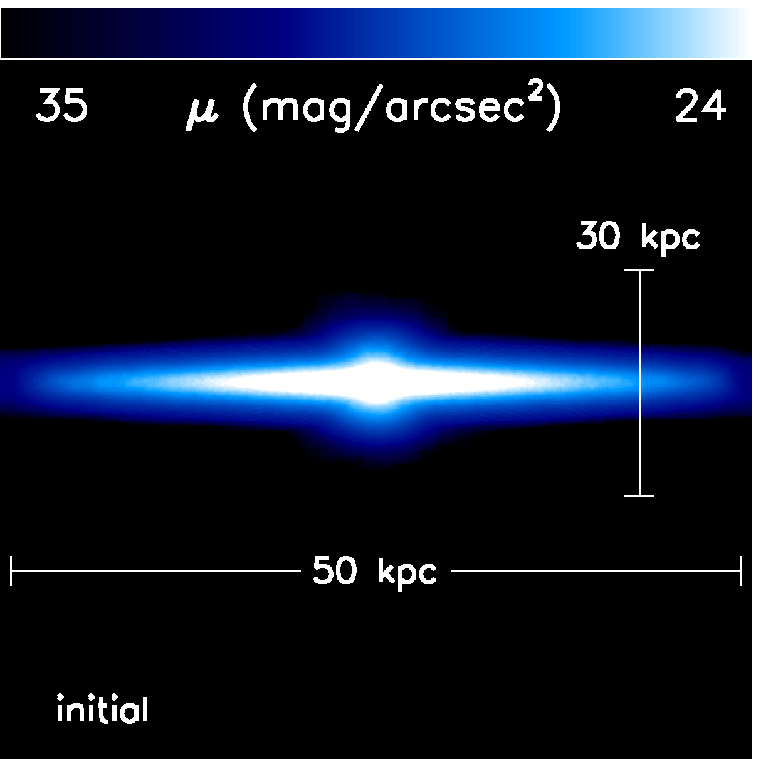} 
\includegraphics[width=2.1in]{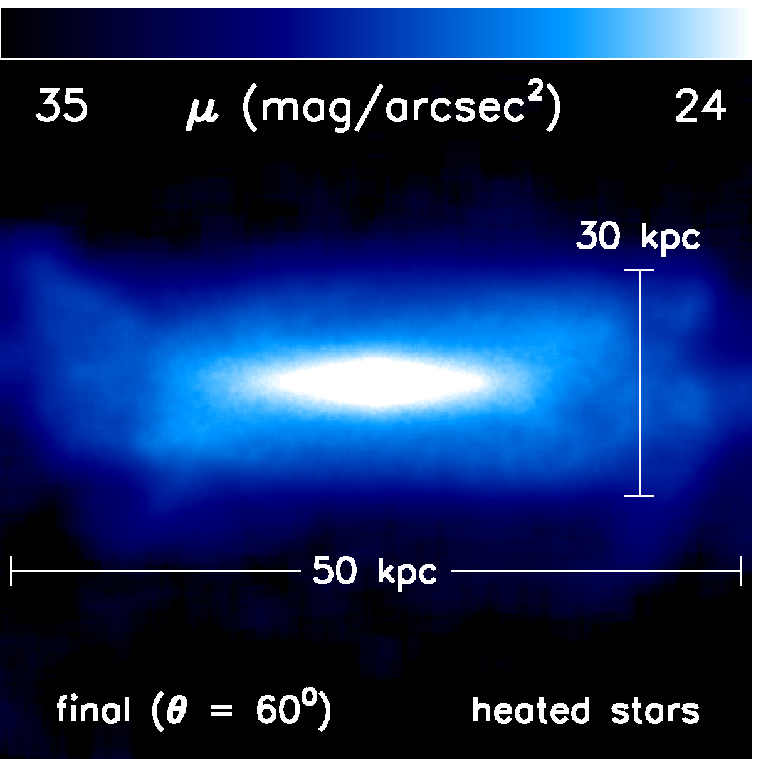} 
\includegraphics[width=2.1in]{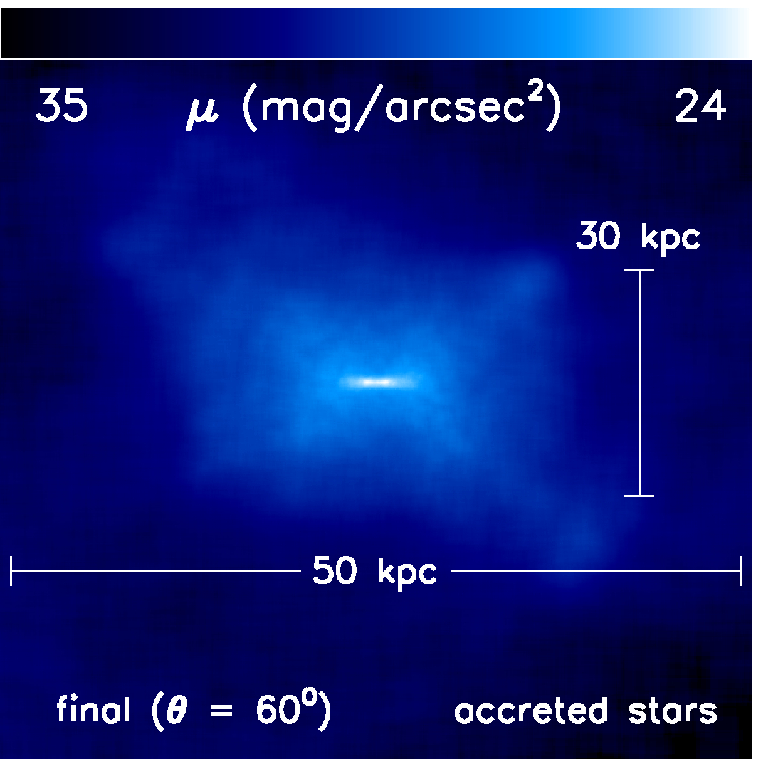}
\caption{Surface brightness maps of our primary system, viewed edge-on: the initial model is visualized in the {\em left} 
panel, while the {\em center} panel shows only stars belonging to the primary galaxy following the prograde infall with 
orbital inclination of $60^{\circ}$, and the {\em right} panel subsamples the accreted stars only.  In all cases, surface 
densities have been converted into surface brightnesses using a stellar-mass-to-light ratio $M_{\star}/L = 3$; note that 
the brightness limits have been chosen in order to accentuate stellar halo substructure at the expense of saturating the 
galactic disc region. }
\label{fig:maps}
\end{figure*}

There are many sources for hot stellar halo material besides direct accretion.  Gas-rich mergers in the early epochs of 
galaxy assembly are likely important for the creation of inner halo material \citep{Brook_04a,Brook_04b}.  Another related 
possibility is that {\em in-situ} stars which form at high redshift ($z \ga 3$) become ejected to large galactocentric radii by 
subsequent major mergers between $2 \la z \la 3$  \citep{Zolotov_etal09}.  In small galaxies, supernovae can affect the 
galactic structure in such a way that primary galaxy stars become liberated into an extended, stellar halo distribution 
\citep{Stinson_09}.  Here, we focus on minor mergers of the type that should be common in the last $\sim 10$ Gyr of 
galactic assembly \citep{Stewart_etal08}.  These accretion events involve satellites massive enough to severely heat the host's 
stellar distribution, but not massive enough to completely eliminate the presence of the primary disc component 
\citep{Kazantzidis_etal08,Read_etal08,Villalobos_Helmi08,Purcell_etal09,Moster_09,Kazantzidis_etal09a}.  

Using a suite of high-resolution collisionless simulations as well as resimulated test cases involving a full treatment of 
hydrodynamical physics, we investigate the process of dynamical stellar ejection induced by the accretion of 
satellites onto a thin galactic disc, and demonstrate that disc stars can become members of the stellar halo, as it is defined 
both spatially and kinematically, via local and global heating modes.  In order to determine the dynamical response of a disc 
to a cosmologically-motivated accretion event, we simulate the interaction between a Galaxy-analogue primary system with 
virial mass $M_{\rm host} \simeq 10^{12} M_{\odot}$ and an infalling satellite subhalo much more massive than the Galactic 
disc itself, $M_{\rm sat} \simeq 10^{11} M_{\odot} \sim 3M_{\rm disc}$.  This mass ratio $M_{\rm sat}$:$M_{\rm host} \simeq$
~1:10 has been shown, via analytic formulations \citep[][see also \citealt{Zentner07}]{Purcell_etal07} as well as cosmological 
simulations \citep{Stewart_etal08,Boylan-Kolchin_09}, to represent the dominant mode of mass delivery into CDM halos, and is therefore of 
immediate concern for the question of disc survivability \citep[as addressed by][]{Purcell_etal09}.  

The majority of the simulations we present here have been discussed elsewhere \citep{Purcell_etal09,Purcell_etal09b,Kazantzidis_etal09a}.  
In \citet{Purcell_etal09} we argued that the resultant discs are likely not good analogues to the Milky Way.  Nevertheless, it very well 
may be that the Milky Way is unusually thin, cold, and old.  The resultant discs are not necessarily inconsistent with the majority of 
external disc galaxies \citep{Moster_09}, which do seem to have thicker scale heights than the Milky Way \citep{YD_06}, though 
observational uncertainties associated with dust-lane obscuration and other effects are important caveats for these considerations.  
Our goal here is to examine a plausible scenario for the emergence of an inner stellar halo component in most galaxies.  They should 
not be compared directly to the Milky Way without some caution (see \S 3.1).  In \S 2, we outline our numerical experiments, following 
with a presentation of our results in \S 3, reserving \S 4 for conclusions and discussion.

\section{Methods}
\label{sec:methods}

We utilize high-resolution, multi-million-particle simulations in order to analyze the response of a thin stellar disc to the infall 
and accretion of a satellite galaxy one-tenth as massive as the host halo itself.    Using fully self-consistent 
numerical models for both the primary and satellite galaxies, as well as initial conditions for accreting subhalos drawn 
from distributions defined by large-scale cosmological simulations, we investigate a range of orbital infall 
inclinations.  Alternative descriptions of our simulations are provided in Purcell et al. (2009a,b); we now review the essential 
features for completeness.

All of our numerical investigations use the multi-stepping, parallel, 
tree $N$-body code PKDGRAV \citep{Stadel2001}, in which we set the gravitational softening length to 
$\epsilon =$ 100 pc and 50 pc for dark matter and stellar particles, respectively.  Each simulation is 
collisionless, with the exception of two experiments in which we ascertain the effect of a full treatment of 
hydrodynamical physics on the ejection of both old and newly-formed disc stars from a primary galaxy with a modest initial 
gas fraction $f_g=0.1$. The latter simulation was performed with the TreeSPH $N$-body code GASOLINE 
\citep{Wadsley_etal04} and is part of a campaign designed to investigate the effect of a dissipative 
component on the dynamical response of thin galactic discs subject to bombardment by halo substructure 
\citep[][in preparation]{Kazantzidis_etal10}.  In these cases, the gravitational softening length for the gas particles was set to 
$\epsilon =$ 50 pc. 

Our primary galaxy components, drawn from the fully self-consistent distribution functions advanced by \citet{Widrow_etal08}, 
represent equilibrium solutions to the coupled collisionless Boltzmann and Poisson equations, and are thus ideally suited 
for simulating the complex dynamics involved in the accretion of a massive satellite galaxy.  The host system consists of a 
massive disc as well as a central bulge following a S\'ersic profile of effective radius $R_e=0.58$~kpc and index $n=1.118$ 
(with mass $M_{\rm bulge}=9.5 \times 10^9 M_{\odot}$ distributed among $5 \times 10^5$ particles); this is the model 
labeled G1 in \citet{Purcell_etal09}.  The stellar disc is initialized with an exponential scale length $R_d=2.84$~kpc and a 
vertical distribution described by a sech$^2$ function with scale height $z_d=0.43$~kpc, and contains $10^6$ particles 
comprising a total disc mass $M_{\rm disc} = 3.6 \times 10^{10} M_{\odot}$.  


\begin{figure}
\begin{center}
\includegraphics[width=3.3in]{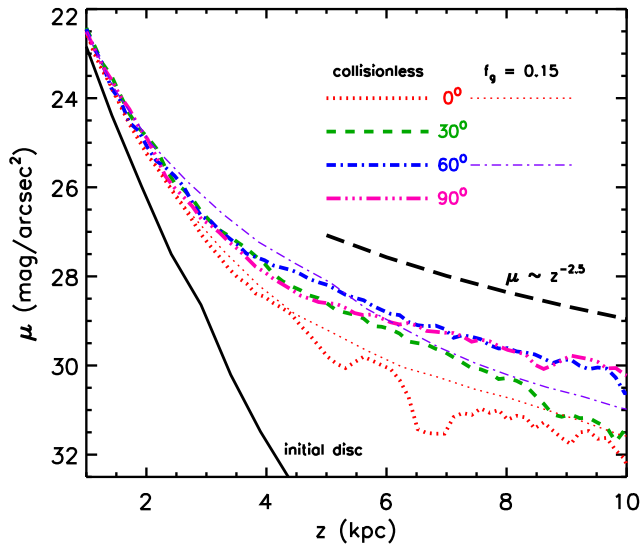}   
\end{center}
\caption{Central surface brightness profiles as measured along the minor axis for each simulation endstate, in both the 
collisionless ({\em thick} lines) and hydrodynamical ({\em thin} lines) regimes, as a function of orbital inclination angle.  
Here, as in Figure~\ref{fig:maps}, we assign a surface brightness $\mu$ based on the stellar-mass-to-light conversion 
factor $M_{\star}/L = 3$.  For reference, at heights $z > 5$~kpc, we plot a profile with arbitrary normalization and 
power-law index $n=-2.5$.}
\label{fig:halo}
\end{figure}

In the hydrodynamical runs, a fraction $f_g=0.15$ of the stellar disc mass is converted into gas, and this gaseous 
component is constructed with the same initial density distribution as the stellar disc; the adopted methodology will 
be described in detail in \citet[][in preparation]{Kazantzidis_etal10}.  We include atomic cooling for a primordial mixture 
of hydrogen and helium, and the star formation algorithm is based on that of \citet{Katz92}, in which gas particles in 
cold and dense Jeans-unstable regions as well as convergent flows spawn star particles at a rate proportional to the local 
dynamical time. In our particular application, gas particles are eligible to form stars if their density exceeds 0.1~cm$^{-3}$ 
and their temperature drops below $1.5\times10^4$~K. Feedback from supernovae is treated using the blast-wave model 
described in \citet{Stinson_etal06}; in this model, the energy deposited by a Type-II supernova into the surrounding gas is 
$4\times10^{50}$~erg. This choice of parameters and numerical techniques produces realistic galaxies in cosmological 
simulations \citep{Governato_etal07}.  Each primary galaxy model is embedded in a dark host halo composed of 
$4 \times 10^6$ particles and arranged in a structure following the canonical NFW density profile of \citet{Navarro_etal96}, 
with scale radius $r_s=14.4$~kpc and virial mass $M_{\rm host} \simeq 10^{12} M_{\odot}$.  We note that our primary 
galaxy model is a Galactic-analogue system chosen from the \citet{Widrow_etal08} set of models with characteristics 
corresponding closely to criteria determined observationally for the Milky Way, and that our precise choice of the initial 
parameter set minimizes the effects of secular evolution such as the formation of a strong central bar, as well as artificial 
heating induced by the interaction of disc particles with more massive halo particles.  

The satellite galaxy simulated in each isolated accretion event is initialized using the self-consistent formalism of \citet{KD95}, 
with which we embed a total stellar mass $M_{\star} = 2.2 \times 10^9 M_{\odot}$ \citep[adopting the value for 
$M_{\star}/M_{\rm sat}$ at $z \sim 0.5$ derived in the number-density matching exercise of][]{Conroy_Wechsler09}, comprised 
of $10^5$ particles populating a spheroidal distribution with S\'ersic index $n \sim 0.5$ \citep[according to the findings of][for 
dSph shape parameters as a function of magnitude in Virgo cluster member systems]{vanZee_etal04}, into a dark subhalo 
composed of $10^6$ particles with a virial mass $10^{11} M_{\odot}$ and having a density structure well-fit by an NFW profile 
with concentration $c_{\rm vir} \simeq 14$ at $z \sim 0.5$.  This value of $c_{\rm vir}$ is consistent with subhalo distributions 
drawn from cosmological simulations \citep[][see also \citealt{Maccio_etal07}]{Bullock_etal01b}.

To investigate the dependence of heating efficiency during a satellite-disc interaction on the orbital parameters of the accreting 
subhalo, we simulate a range of inclination angles for each primary galaxy model: $\theta = 0^{\circ}, 30^{\circ}, 60^{\circ}, 
\mathrm{and}~90^{\circ}$ (where $\theta$ is defined as the angle between the angular momentum axes of the galactic disc and 
the infall orbit).  In each experiment, the satellite galaxy is initially placed at a large distance $r \simeq 120$ kpc from the center 
of the host halo, to minimize the shock imparted to the primary galaxy particles by the sudden appearance of this new potential 
well.  The initial velocity vector for the subhalo is motivated by cosmological analysis of substructure accretion, in which the 
distributions of radial and tangential velocity components for satellite systems ($v_r$ and $v_t$) peak respectively at 90\% and 
60\% of the host halo's virial velocity \citep{Benson2005,Khochfar_Burkert06}; therefore, our subhalo vectors have $v_r = 116$ 
km/s and $v_t = 77$ km/s.  

All fiducial simulations involve orbits that are prograde with respect to the primary galaxy's rotation; we also perform a retrograde 
simulation for the inclination angle $\theta = 60^{\circ}$ in order to assess the degree to which this case decreases the efficiency 
of disc heating while increasing the amount by which the primary galaxy is tilted away from its initial angle of repose, as expected 
according to previous experiments involving satellite-disc interactions \citep{Velazquez_White99,Kazantzidis_etal09a}.  Each 
prograde simulation is allowed to evolve for a total of 5 Gyr, by which time the subhalo has fully coalesced with the host halo and 
the heating process is virtually complete, leaving the galaxy in a quasi-steady state; the elapsed timescale for the retrograde model 
is slightly longer at 7 Gyr, since dynamical friction is less efficient in this case. Lastly, the hydrodynamical simulations employ 
prograde orbits with inclination angles for the infalling satellite of $\theta = 0^{\circ}~\mathrm{and}~60^{\circ}$.


\begin{figure*}
\begin{center}
\includegraphics[width=5.5in]{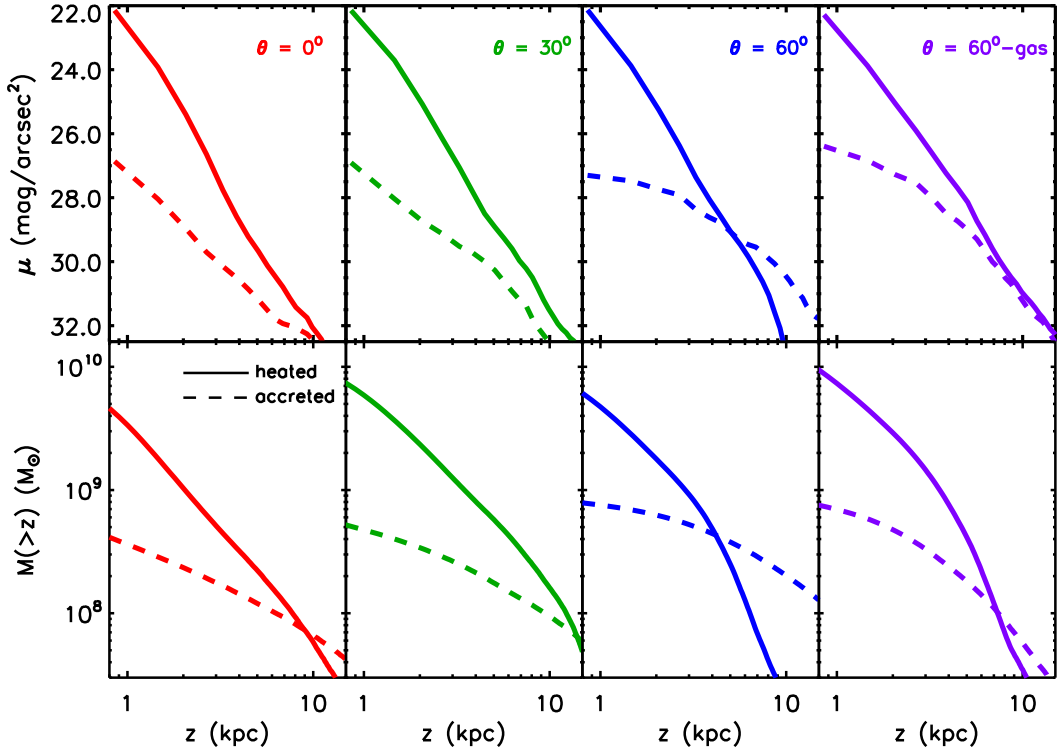} 
\end{center}  
\caption{Minor-axis surface brightness $\mu$ ({\em upper} panels) and cumulative stellar mass $M(>z)$ ({\em lower} panels) contributed by 
heated disc stars ({\em solid} lines) and accreted satellite stars ({\em dashed} lines) at height $z$ from the galactic plane.}
\label{fig:eightpanel}
\end{figure*}

 \section{Results}
\label{sec:results}

Regardless of orbital inclination, each fiducial accretion event ejects a significant portion of stellar mass from the disc, as shown 
qualitatively in the visualizations of Figure~\ref{fig:maps}.  This material reaches heights above the galactic plane consistent with 
the extended faint spheroids observed in many nearby massive disc galaxies, most notably in the diffuse inner halo of M31 
\citep{Guha_etal05,Koch_etal08}.  The disc stars still located at the mid-plane of our simulated disc remnants are much dynamically 
hotter than the initially cold and thin stellar system, yielding a distribution of velocities consistent with a two-component disc as well 
as a non-trivial fraction of stars that have kinematics similar to those found in the stellar halo of the Milky Way. 

Figure~\ref{fig:halo} presents the minor-axis surface-brightness profiles of the initial disc galaxy (solid, black) and the resultant discs
after each encounter, as measured along the minor axis extending from the galactic center\footnote{We choose the central region since 
it is the brightest and thus most well-constrained area in observation of external galaxies.  The minor-axis profile does not change in 
slope significantly as galactocentric radius increases, although the disc outskirts at $R \ga 10$~kpc show pronounced flaring at faint 
surface brightnesses.}.  We see that the post-merger disc 
surface-brightness profile generally begins to show a distinct transition to a diffuse power-law component at 
$\mu \sim 28$~mag/arcsec$^2$ (for a stellar-mass-to-light ratio $M_{\star}/L = 3$).   Shown for comparison is a power-law slope 
of $\mu \propto z^{-2.5}$, which is indicative of that expected for a stellar halo.  As measured and reported by \citet{Purcell_etal09} 
using a two-component sech$^2$ fitted profile, the scale height of the remaining thin-disc component varies along the range 
$z_{\rm d} \sim 1-2$~kpc according to the satellite's orbital inclination.  We note that the thicker component of the fit invariably has 
a very large scale height ($z_{\rm diffuse} \sim 4-7$~kpc), reflecting the large amount of energy imparted through global and direct 
heating modes during the accretion event.

Generally, our experiments result in a significant portion of the initial disc's stellar mass being ejected by the satellite-disc interaction 
into a diffuse spheroidal component at heights greater than $z \ga 5z_{\rm d} \sim 7$~kpc.  Figure~\ref{fig:eightpanel} shows the 
surface brightness profiles (as in Figure~\ref{fig:halo}) as well as the cumulative stellar mass found above a given height $z$ from the 
disc plane, with both quantities having been split into the components contributed by the heated disc stars and the accreted satellite 
stars.  We see that low-latitude accretion events ($\theta = 0^{\circ},30^{\circ}$) heat the discs so efficiently that a stellar mass 
equivalent to roughly $1\%$ of the initial disc mass, $M_{\star}^{\rm halo} \sim 2-3 \times 10^8 M_{\odot}$, is ejected to heights 
larger than 7 kpc.  For high-latitude accretion events ($\theta = 60^{\circ},90^{\circ}$), tidally-stripped stars from the accreted satellite 
dominate the population, and disc-star ejection contributes $M_{\star}^{\rm halo} \sim 10^8 M_{\odot}$.  Closer to the galactic 
mid-plane, the very thick stellar disc becomes the prevalent factor regardless of inclination angle; approximately $10^9 M_{\odot}$ 
in initial-disc stars alone are ejected to a height larger than $z \sim 4$~kpc, across the full radial extent of the disc.  


\begin{figure*}
\begin{center}
\includegraphics[width=6in]{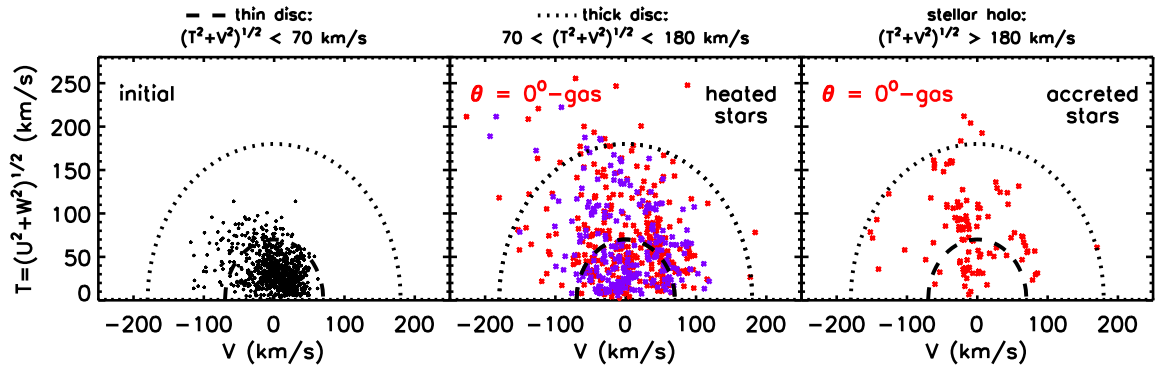}  
\end{center}
\caption{The solar-neighborhood Toomre energy diagram for the initial model ({\em left} panel) and the endstate system including 
heated disc stars ({\em center} panel) and accreted stars ({\em right} panel) following the satellite galaxy's prograde infall at an orbital 
inclination of $0^{\circ}$ onto a primary galaxy with an initial gas fraction $f_g = 0.15$; the quantity $T$ is the quadratic sum of the 
radial and vertical space velocity components $U=v_{R}$ and $W=v_{z}$, and $V$ is the rotational velocity $v_{\phi}$.  Disc stars are 
subsampled from the galactic discs in a very thin annulus (100 pc in width and 300 pc in height) centered vertically on the plane of 
the disc and sweeping through the circumference defined by the solar radius $R_{\odot} = 8$ kpc; the annulus width and height are 
each tripled for sampling of the accreted stars, to minimize numerical noise.  The origin of the x-axis is set in each case by the local 
standard of rest, defined as the mean rotational velocity $V$ of the stellar subsample.  We define stars inside the enclosing {\em dashed} 
line with a radius of 70 km/s to be kinematic members of the thin disc, while stars outside this region and enclosed by the {\em dotted} 
line with a radius of 180 km/s are members of the thick disc \citep[as in][]{Nissen_Schuster09}; subsampled stars outside this region are 
kinematic members of the stellar halo despite their spatial location near the surface brightness peak at the disc's mid-plane.  In this 
hydrodynamical test case, {\em red} points mark original disc stars and {\em purple} points denote stars formed during the simulation's 
evolution.}
\label{fig:toomre}
\end{figure*}

Despite some thin-disc regrowth via star formation in the hydrodynamical simulations, the vertical disc thickening is even more pronounced 
in those cases than in the collisionless experiments, possibly because the initial stellar disc is less massive and thus more susceptible to the 
strong global resonance modes induced by the high-latitude accretion event.  We note that in the hydrodynamical regime, the increase in 
friction accelerates the subhalo infall and thus results in less material stripped from the satellite galaxy at large heights above the primary 
galaxy plane; these accretion events are typically complete a few hundred Myr before their collisionless counterparts.

\subsection{Applicability to the Milky Way and M31}
\label{subsec}

Recent surveys indicate that the Galactic stellar halo may somewhat possess a dual nature similar to that observed in M31, being comprised of 
an outer spheroid of metal-poor stars having a slightly retrograde rotation with respect to the Milky Way disc, and an inner component with a 
flatter axial ratio and a mild prograde rotation, as well as stars more metal-rich than the classic extended halo by a factor of three 
\citep{Chiba_Beers00,Carollo_etal07,Kinman_etal07}.  Although most of the Galactic halo is composed of an ancient stellar population, younger 
and richer stars have been observed there for decades \citep{Greenstein_Sargent74,Siegel_etal09}; a small fraction of these objects have peculiar 
velocities such that their calculated flight times from the disc are longer than their apparent lifetimes, and thus some debate exists over whether 
these stars formed {\em in situ} in the low-density gas of the stellar halo \citep{Keenan92}, or are possibly undergoing binary rejuvenation 
\citep{Perets09}, or were ejected from the disc via dynamical interactions either internal to a multiple star system \citep{Blaauw61,Gualandris_etal04} 
or involving the massive black hole at the center of the Milky Way \citep{Hills88,Yu_Tremaine03}.  On a larger interaction scale, the simulations of 
\citet{Abadi_etal09} have shown that disrupting dwarf galaxies may leave behind stars with velocities similar to or exceeding the escape speed of 
the primary system, which may cause some degree of angular collimation and travel-time commonality in a hypervelocity population, as observed 
in the fast-moving Galactic halo stars clustered in the Leo constellation.

The majority of runaway disc stars, however, are not these hypervelocity objects but instead have moderately hot kinematics similar to those of 
the Milky Way's classical halo, although they tend to retain significant rotational velocity as well \citep{Martin2006}, and can even be found far 
above the Galactic plane \citep[][see also \citealt{Allen_Kinman04} and references therein]{Ramspeck_etal01}.  In a spectroscopic survey of stars 
belonging to both the thick disc and the stellar halo, \citet{Nissen_Schuster09} report a number of stars with halo-like speeds, and unexpectedly 
high $\alpha$-enrichment, surmising that this could imply either a dissipational collapse mechanism in the halo \citep[as in][]{Gratton_etal03} 
or the signature of an accretion event involving a large satellite galaxy.  

In the context of the Milky Way, we can distinguish dynamically between stars in the thin/thick 
disc structure and stars that have been heated into orbits similar to those found in the stellar halo by employing the analysis of Toomre energy space, 
in which the radial ($U$) and vertical ($W$) space velocities are summed in quadrature ($T=[U^2+W^2]^{1/2}$) and diagrammed against the rotational 
velocity $V$ \citep[][see \citealt{Carollo_etal07} for a discussion of the solar neighborhood as viewed in this parameter space]{Sandage_Fouts1987}.  
\begin{figure*}
\begin{center}
\includegraphics[width=6.6in]{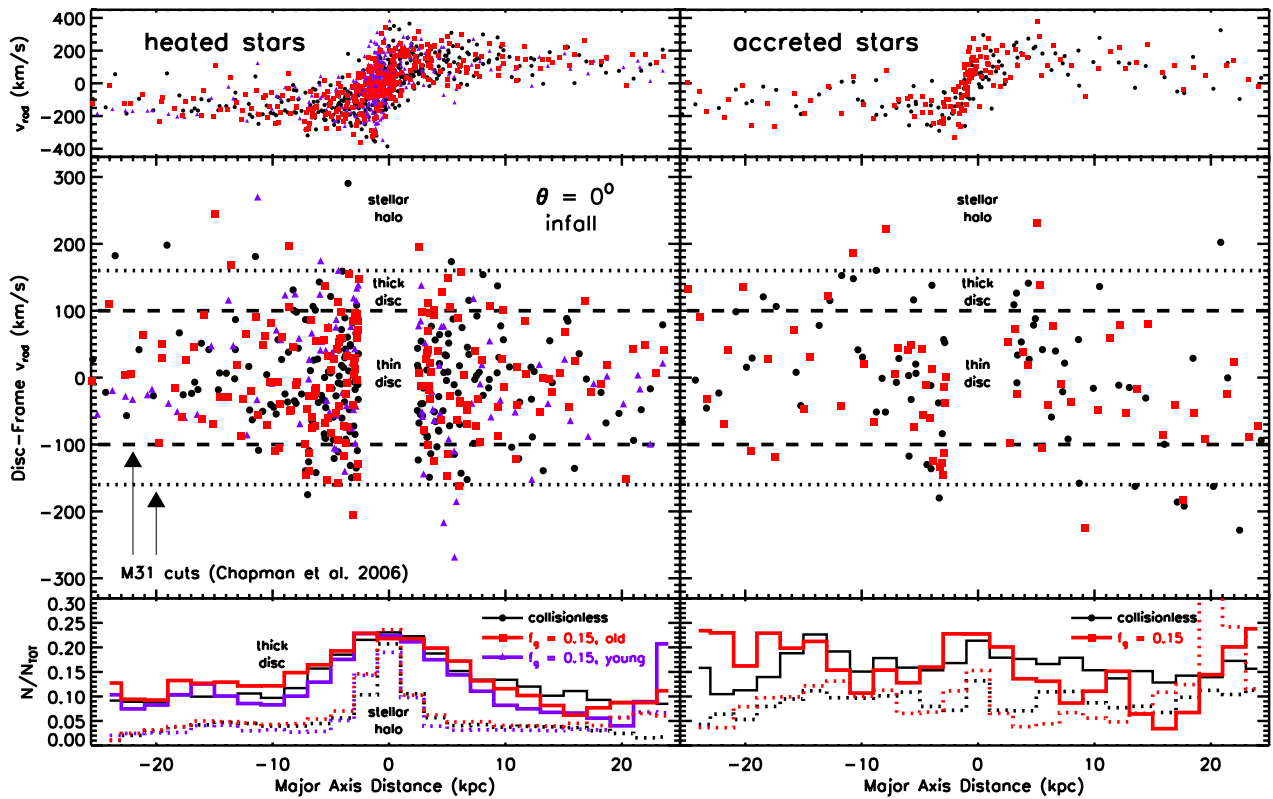}   
\end{center}
\caption{{\em Upper:} Stellar radial velocities (from the edge-on perspective) as a function of distance along the major axis, for subsamples of the collisionless 
({\em black circles}) and hydrodynamical 
({\em red squares} for old stars and {\em purple triangles} for newly-formed stars) events involving planar infall ($\theta = 0^{\circ}$).  {\em Center:} Velocities have been 
corrected into the kinematic frame of the disc, {\em i.e.}, the mean local value of the rotation curve has been subtracted from each point for ease of comparison to the 
analysis of M31 by \citet{Chapman_etal06} (see \S\ref{subsec}), in which the authors define boundaries in this phase space for each kinematic component.  Note that the 
inner few kpc are excised in these panels to relieve overcrowding in the region where the mean radial velocity decreases sharply, and that these panels represent an 
appropriately-sized subsample of the particle distribution.  {\em Lower:} The relative fraction of stars in the thick disc ({\em solid}) and stellar halo ({\em dotted}) 
components defined by their radial velocity ranges, as a function of distance along the major axis; the innermost disc is restored in this panel and the entire particle 
distribution is represented.}
\label{fig:radial}
\end{figure*}
In Figure~\ref{fig:toomre}, we extract disc stars from a thin annulus sweeping through the solar locality and centered on the mid-plane, with a radial width 
of $100$~pc and thickness of $300$~pc, and plot the kinematic properties of this population.  Despite being located spatially near the mid-plane of the 
post-accretion thin-thick disc structure, the number of stars with thick-disc kinematics has increased significantly when compared to the dynamical 
composition of the initial galaxy models, regardless of the satellite galaxy's orbital inclination during infall.  Moreover, non-trivial numbers of disc stars 
near the plane at the solar radius have been heated so dramatically by the accretion event that their three-dimensional velocities are as high as those 
observed in the Galactic stellar halo.  By comparison, the kinematics of the accreted satellite stars are consistent with a mixed population of thick-disc and 
stellar-halo stars, as determined in large part by the rotational speed of the accreted material, which lags the primary disc by an amount that correlates with 
increasing infall inclination angle \citep[see][for an extended discussion of this behavior as it pertains to the dynamical relationship between accreted stars 
and accreted dark matter]{Purcell_etal09b}.

Though the results presented in Figure~\ref{fig:toomre} are intriguing, it appears highly unlikely that the Milky Way has recently suffered such a dramatic accretion event 
as those we fiducially investigate here; not only is the Galactic disc far 
too cold and thin to have undergone much global heating in the last several Gyr \citep{Purcell_etal09}, it is also quite deficient in stellar mass and angular momentum 
when compared to a sample of local spiral galaxies \citep{Hammer_etal07}.  These constraints indicate that the accretion history of the Milky Way has been substantially 
more quiescent than would be expected for a typical halo of virial mass $M_{\rm host} \sim 10^{12} M_{\odot}$ in the $\Lambda$CDM cosmology.  On the other hand, 
it is plausible to suggest that minor events such as the ongoing accretion and disruption of the Sagittarius dwarf galaxy may also be capable of heating some small 
portion of disc stars to ejection, especially in the outer disc where the disc's self-gravity is less powerful; we may safely assume that the Galaxy has undergone some 
number of accretions in this regime, these events being truly ubiquitous in a cosmological context.

At least one member of the Local Group may represent more nearly the cosmological norm; the inner spheroid of M31 has been chemically identified as having a stellar 
population that is largely of intermediate age and as metal-rich as that galaxy's thick disc \citep{Durrell_etal04}.  It may well be that M31, having had a more active recent 
accretion history than the Milky Way \citep[as surmised by][based on density and metallicity variations in halo substructure]{Ferguson_etal02,Gilbert_09b}, has a stellar halo 
that has been significantly contaminated by stars originally belonging to the primary disc.  Some indications do exist that this is the case; uniform and young stellar populations 
have been shown to exhibit strong rotation out to projected radius $R_{\rm proj} > 40$~kpc \citep[][see also \citealt{Ferguson_etal05}]{Richardson_etal08}.  In addition, the 
Giant Southern Stream and associated stellar features are clearly the tidal remnants of an accretion event involving a fairly massive satellite galaxy; estimates of the progenitor's 
properties suggest that it was a large system traveling on a highly-eccentric orbit of nearly co-planar orientation with respect to the M31 disc, and the surface brightness of the 
tidal stream strongly implies that the event occurred within the past few Gyr \citep{Ibata_etal04,Fardal_etal06,Font_etal06}.  Our result for the simulated infall with orbital 
inclination $\theta = 0^{\circ}$ predicts a substantial amount of disc-star ejection associated with such an accretion event, an expectation that is consistent with the relatively 
bright inner halo of M31.

Along these lines, in Figure~\ref{fig:radial} we present an analogous consideration to that found in the study of metal-poor M31 halo stars by 
\citet[][hereafter C06]{Chapman_etal06}, in which the authors utilize 
the kinematic model of \citet{Ibata_etal05} in order to correct observed heliocentric radial velocities for the systemic and rotational motion of the M31 disc as well as the 
disc's inclination.  Following C06, we present line-of-sight velocities across the major axis for our $\theta=0^{\circ}$ runs from the edge-on perspective (upper 
panels) and subtract the rotational motion (middle panels).  C06 used velocity cuts to kinematically represent the thin disc ($|v_{\rm rad}| < 100$~km/s), 
thick disc ($100 < |v_{\rm rad}| < 160$~km/s), and stellar halo ($v_{\rm rad} > 160$~km/s) of M31; we adopt the same cuts here for comparison, including only heated 
stars in the left panels and only accreted stars in the right panels.  We obtain significant relative fractions in each component across the full length of the major axis, quite 
similar to the results of C06 for M31.  Across the full length of the extended discÕs major axis, approximately $\sim 10\%$ of heated disc stars have radial velocities 
consistent with the thick disc criterion set by C06 for M31, and $\sim 5\%$ are consistent with the kinematic stellar halo definition.  In the inner disc, where the rotation 
curve drops sharply and the bulge dominates the stellar density (although the bulge stars have been removed from this figure), the dynamical contamination is even more 
pronounced: the thick disc and stellar halo represent $\sim 15-25\%$ each of the radial velocity population. Stellar material deposited by the infalling satellite also populates 
these regions; around $\sim 10-20\%$ of these stars fall outside the thin-disc range in radial velocity.

 \section{Discussion}
\label{sec:discuss}

Observations of diffuse light in the inner halos of external galaxies have tended to inspire efforts to reconcile the stellar mass of the spheroid with the properties 
of an inferred progenitor satellite galaxy.  However, our experiments clearly demonstrate that an inner halo component may be primarily composed of stars 
ejected to large heights above the galaxy plane during the disc heating process.  In our post-accretion remnant systems, we obtain surface brightness profiles 
typified by a power-law index $n=-2.5$ along the minor axis as shown in Figure~\ref{fig:halo}, only slightly flatter between $5$ and $10$~kpc than the projected 
profile found to be a general feature by \citet{Zibetti_etal04} in their image-stacking exercise involving over a thousand SDSS galaxies.  In each of our simulation 
endstates, ejected material originally belonging to the primary stellar disc accounts for $\sim 40-80\%$ of the luminosity found above $5$~kpc from the plane, 
with the accreted subhalo's stars being distributed over a much larger volume and thus contributing minimally to the surface brightness of the inner halo.  

This morphological contamination of the inner halo occurs simultaneously with the energizing process that also significantly contributes stellar material to the 
kinematically-defined thick disc and stellar halo.  Just as in models of stellar halo formation, the scenario of thick disc assembly most likely involves multiple 
mechanisms, although there is active debate regarding the order of importance in these phenomena.  The wide range of kinematic behavior displayed by low-mass thick disc 
galaxies may indicate, especially for cases in which signatures of counter-rotation appear, that these systems form primarily via the accretion of stripped stars during 
a minor merger, or from star formation associated with such a merger \citep{Yoachim_Dalcanton08}.  

On the other hand, there is solid photometric evidence that more massive disc galaxies also contain thick components that are simply too faint to discern kinematically, 
especially in the absence of significant rotational lag in any particular stellar population.  This co-rotation is a particular hallmark of low-latitude subhalo accretion, 
since in this case the satellite's stars will kinematically blend more efficiently into the host's rotation, and such uniform dynamics are also a general feature of thick discs 
formed largely by merger-related heating processes, which can achieve line-of-sight velocity dispersions similar to those found in surveys of these systems.  In the case 
of the Milky Way, the thick component's faintness relative to the thin disc, as well as its measurably lagging rotation, likely indicate that it is not a product of heating that 
has occurred in the last several Gyr; observations have yet to determine whether the kinematic structure of the thin/thick Galactic disc is typical among systems with similar 
morphology and stellar content. 

Similarly, it remains to be shown self-consistently that a cold and thin Galaxy analogue can arise and persist to $z=0$ after the thick disc's formation at early times during 
the process of heating the proto-Galactic disc.  However, we may still make a simple argument in favor of disc heating being largely responsible for the complex kinematic 
distributions in today's solar neighborhood.  As alluded to above, the total mass of the disc under assault is an integral factor in any accretion event; according to first-order 
expectations, the strength of the stellar self-gravity will largely determine the extent to which any deposited orbital energy (or excited global resonance) heats the system.  As 
an approximation, we would therefore expect more severe dynamical and morphological damage to be inflicted by an accretion event involving a lighter disc, such as the 
Galactic progenitor at intermediate redshift, leaving behind some fast-moving relics of this era in nearby stellar populations at the present day.  A more detailed exploration 
regarding the role of disc mass in the accretion-heating process can be found in \citet[][in preparation]{Kazantzidis_etal10}.  We also note here that although previous work 
has indicated that multiple minor mergers are unlikely to cause more significantly more structural and dynamical damage than that inflicted by the most massive satellite 
alone \citep{Kazantzidis_etal08}, it may be that mixing processes and resonance modes are enhanced by these multiple infalls; we defer pursuit and discussion of these 
issues for future work.

It appears that the formation of inner galactic halos may be heavily enhanced by the destructive heating undergone by a stellar disc during the infall of a large satellite galaxy.  
As we improve our understanding of stellar halo composition in nearby disc galaxies, we may begin to assess the degree to which these diffuse spheroids are alloyed by 
intermediate-age stars that are unlikely to have been part of the primeval halo constructed long ago from the debris of faint and metal-poor dwarf satellites.  Future Galactic 
surveys, such as SEGUE-2 and APOGEE of the third iteration in the Sloane Digital Sky Survey \citep[SDSS-III;][]{Weinberg_etal07}, will map the outer disc and stellar halo of the 
Milky Way to an unprecedented degree of accuracy, and M31 observations will continue to refine our understanding of that galaxy's complex inner structure.  These efforts will 
exquisitely define the kinematics, density structures, and chemical compositions of the various populations in each component, and should easily be able to distinguish the relative 
properties of ancient, metal-poor halo members as compared to younger, more metal-rich stars that could have been heated to ejection from the disc during a massive 
accretion event.  

$\;$

We would like to thank Larry Widrow and John Dubinski for kindly making available 
the software used to set up the initial galaxy models, and David Weinberg for useful 
discussions.  We also thank Nicolas Martin for very helpful suggestions during the 
referee process, which improved the manuscript greatly.  
CWP and JSB are supported by National Science Foundation (NSF) 
grants AST-0607377 and AST-0507816, and the Center for 
Cosmology at UC Irvine.  SK is supported by the Center for Cosmology and 
Astro-Particle Physics at The Ohio State University. The numerical simulations 
were performed on the IA-64 cluster at the San Diego Supercomputing Center, 
with ancillary experiments performed on the GreenPlanet cluster at UC Irvine.  
This work was also supported in part by an allocation of computing time from 
the Ohio Supercomputer Center (http://www.osc.edu/).

\end{document}